\documentclass[aps,prl,twocolumn,groupedaddress,showpacs]{revtex4}
\usepackage{graphicx}
\usepackage[dvips]{epsfig}
\usepackage{dcolumn}
\usepackage{subfigure}%
\usepackage{bm}
\usepackage{amssymb}
\usepackage{amsmath}

\begin{document}

\title{Reply to the revised Comment [PRL 102, 139601; arXiv:0810.4791] on ''Dynamic Scaling of Non-Euclidean Interfaces''}

\author{Carlos Escudero}

\affiliation{Instituto de Ciencias Matem\'{a}ticas, Consejo Superior
de Investigaciones Cient\'{\i}ficas, C/ Serrano 123, 28006 Madrid,
Spain}

\begin{abstract}

\end{abstract}


\maketitle

The comment by Krug on Letter \cite{escudero} is twofold: firstly
he argues that our result enters in contradiction with much of the
previous work on surface growth; secondly he claims that the
correlations found in \cite{escudero} are a straightforward
consequence of standard dynamic scaling. Both arguments are
incorrect as will be shown in the following. On one hand, the
mathematical results present in \cite{escudero} are correct, they
do not contradict any of the previous theoretical work, and they
indeed describe the interface dynamics in the long time and large
spatial scale. On the other hand, the correlations of the radial
interfaces calculated in \cite{escudero} cannot be obtained as a
particular limit of the Family-Vicsek ansatz, contrarily to what
was pointed out in \cite{krug}.

First, we consider Krug's assertion on the contradiction of our
calculation with previous work. Letter \cite{escudero} does not
mathematically contradict any of the previous theoretical work. To
compare two results they have to refer to the same system, but
also this system have to be studied in the same limit in both. In
\cite{escudero} we consider the interface dynamics in the long
time large scale limit. This same limit has not been considered in
the previous works cited by Krug, so any direct comparison is not
possible. One can illustrate this point with the calculation by
Singha \cite{singha}. Singha studied the two times one point
correlation function $\langle \rho_n(t)\rho_{-n}(t') \rangle$,
while we studied the one time two points correlation $\langle
\rho_n(t)\rho_m(t) \rangle$: this difference becomes fundamental
when the long time asymptotics is considered. Singha found the
result $C_S(\theta;t,t') \sim \sqrt{\min(t,t')}$, and we found the
result $C_E(\theta,\theta',t) \sim \mathrm{ln}(t) \,
\delta(\theta-\theta')$. This is not a contradiction, as two
different physical limits are being examined. One can see in the
original calculation, Eqs. (19) and (20) in \cite{escudero}, that
the limit considered implies Fourier modes $n^2 \ll t$, what
implies in turn that we are focusing on angular scales $\Theta \gg
1/\sqrt{t}$, and we are excluding the strictly local scale studied
by Singha. A similar argument applies to the other references
cited by Krug.

The second point raised by Krug concerns the connection of the
correlations computed in \cite{escudero} with the Family-Vicsek
ansatz that describes the scaling behavior of planar stochastic
growth equations. According to him, these correlations are a
direct consequence of the mentioned ansatz, this is, the
two-points correlation function is of the Family-Vicsek form
\begin{equation}
C(\theta,\theta',t) \approx t^{2\beta}
\mathcal{C}(|\theta-\theta'|t^{1-1/z}).
\end{equation}
Then, the long time limit {\it as taken in} \cite{krug} yields
\begin{equation}
\label{longtime}
\lim_{t \to \infty} C(\theta,\theta',t) \sim
t^{2\beta -1 +1/z}\delta(\theta - \theta')= t^{(1-d)/z}
\delta(\theta - \theta'),
\end{equation}
{\it according to} \cite{krug}, where the relation $2 \beta = 1-
d/z$ for linear growth equations has been employed in the last
equality. For $d=1$ we found a prefactor $t^0$, what is claimed in
\cite{krug} to be the explanation of the logarithmic prefactor in
\cite{escudero}. This argument is based on two erroneous facts.
First, the long time limit (\ref{longtime}) is incorrect, as the
prefactor necessary to build the Dirac delta function has been
forgotten. The correct calculation would be
\begin{eqnarray}
\label{longtime2} &C(\theta,\theta',t)& \approx t^{2\beta}
\mathcal{C}(|\theta-\theta'|t^{1-1/z})= \\
&t^{2\beta} t^{d/z-d}& \left[ \frac{1}{t^{d/z-d}}
\mathcal{C}(|\theta-\theta'|t^{1-1/z}) \right] \sim t^{1-d}
\delta(\theta - \theta'), \nonumber
\end{eqnarray}
where the long time limit has been taken in the last step.
Prefactors in (\ref{longtime}) and (\ref{longtime2}) are
different, and it is evident that result (\ref{longtime}) is
incorrect for two reasons. First, (\ref{longtime}) is not
compatible with the random deposition correlation $C_{rd} \sim t
\delta(x-x')$ obtained from the Family-Vicsek ansatz, and it does
not coincide with the radial correlations found for $d>1$ in
\cite{escudero} and \cite{escudero2}. Derivation (\ref{longtime2})
is compatible with the random deposition correlation and with
previously calculated radial correlations {\it if dilution is
taken into account}, see \cite{escudero3}. And this is precisely
the second mistake in \cite{krug}. Radial correlations for which
dilution has not been considered do not reduce to either the
incorrect (\ref{longtime}) or correct (\ref{longtime2}) forms. One
can see that, for $d>1$, the prefactor becomes constant and not a
power law of time \cite{escudero,escudero2}. And so, {\it radial
correlations cannot be deduced from the Family-Vicsek ansatz}. On
the other hand, if one takes into account dilution
\cite{escudero3}, radial correlations reduce to the Family-Vicsek
form (\ref{longtime2}) (but not to (\ref{longtime})). The physical
reason is that the stochastic growth equations considered in
\cite{escudero,krug,singha,escudero2} develop memory with respect
to the initial condition, and this memory effect is not captured
by the Family-Vicsek ansatz. Dilution erases this memory, and so
its inclusion implies the recovery of the Family-Vicsek ansatz
\cite{escudero3}. As a side note let us mention that the claim in
\cite{krug} specifying that a prefactor $t^0$ in (\ref{longtime})
(or (\ref{longtime2})) is compatible with the logarithmic
prefactor in \cite{escudero} is incorrect too. Although these
prefactors are compatible from a dimensional analysis viewpoint,
the explicit calculation of the two-points correlation function
shows that this prefactor is constant and not logarithmic when
$d=1$ \cite{escudero3}. Again, the logarithmic prefactor is a
memory effect that cannot be explained using the Family-Vicsek
ansatz.

In summary, the radial correlations calculated in \cite{escudero}
cannot be deduced from the Family-Vicsek ansatz, what implies a
different type of scaling. In consequence, reconsidering those
experimental works in which radial interface profiles were
analyzed applying planar concepts without any justification might
be in order.

\end{document}